# A P2P Context Lookup Service for Multiple Smart Spaces

Tao Gu [1,2], Hung Keng Pung [2], Daqing Zhang [1]

[1] Institute for Infocomm Research, 21 Heng Mui Keng Terrace, Singapore

[2] National University of Singapore, 3 Science Drive 2, Singapore

tgu@i2r.a-star.edu.sg; punghk@comp.nus.edu.sg; daqing@i2r.a-star.edu.sg

**Abstract**

Context information has emerged as an important resource to enable autonomy and flexibility of pervasive applications. The widespread use of context information necessitates efficient wide-area lookup services. In this paper, we present the design and implementation of a peer-to-peer context lookup system to support context-aware applications over multiple smart spaces. Our system provides a distributed repository for context storage, and a semantic peer-to-peer network for context lookup. Collaborative context-aware applications that utilize different context information in multiple smart spaces can be easily built by invoking a pull or push service provided by our system. We outline the design and implementation of our system, and validate our system through the development of cross-domain applications.

## 1. Introduction

The recent convergence of pervasive computing and context-aware computing has seen a considerable rise in interest in various context-aware applications. These applications exploit various aspects of the contextual environment to offer services, present information, tailor application behavior and trigger adaptation, based to the changing context.

Many existing context-aware applications such as Active Badges [1] and ParcTab [2] have demonstrated the benefits of utilizing context information in provisioning context-awareness to various applications. However, these systems have also shown that it is difficult to design and develop robust context-aware applications. There are a number of issues that must be resolved, such as handling diversity of context data, dealing with context acquisition, and maintaining system interoperability, etc. Among those issues, context lookup is an important issue because the primary goal of context-aware computing is to acquire and utilize context information to provide services that are appropriate to people, place, time, events, etc [3].

A common approach is to use a centralized search engine to store context data and resolve lookup queries. It works well in a single smart space. Since there is only a small data set in a single domain, this approach can provide fast responses to a query and it is relatively easy to update data or index. However, the centralized approach has limitations such as a single processing bottleneck and single point of failure. More importantly, it may not scale up well for collaborative context-aware applications that require to access and utilize different sets of context data over multiple smart spaces. It is also difficult to keep data up-to-date due to the dynamicity of context data. Emerging Peer-to-Peer (P2P) approaches have been proposed to overcome some of these obstacles, and provide potential solutions for building non-centralized context lookup systems in multiple smart spaces.

This paper presents the design and implementation of a P2P context lookup system to support context-aware applications over multiple smart spaces. In this system, context data are distributely stored in various context producers over different domains. Each context producer is only responsible for managing its local data that may be acquired from the sensors attached. Context data are represented by RDF (Resource Description Framework) statements according to ontologies. To facilitate users and applications an efficient context lookup service in multiple smart spaces, we design and implement a semantic P2P network in which RDF-based context data are organized and retrieved according to their semantics. In this network, context producers are arranged in such a way that those with semantically similar data are grouped together so that a context query can be routed efficiently. Both pull and push services are provided to meet the needs of various applications.

In summary, this paper makes the following contributions:

- As far as we know, we are the first to design and implement a P2P context lookup system for multiple smart spaces in pervasive computing.

- We design various techniques to meet the requirements of programmability, scalability and dynamicity in context-aware systems, such as sensor wrappers for acquiring sensor data easily, a P2P context repository for storing context data, one-dimensional ring space for facilitating efficient query routing, and a push service for notifying context consumers about context changes quickly.

- We demonstrate the practicality of our system by developing several real-life applications in multiple smart spaces. Our experiences show that the development process can be greatly simplified.

## 2. ContextPeers Architecture

In our system, context producers provide various context data for sharing; whereas context consumers such as users or applications obtain context data by submitting their queries to and receiving results from the network. The system consists of many context producer peers called ContextPeers that are self-organized into a semantic P2P network – SCS [4]. ContextPeers exploits SCS as the underlying network layer and extends it with RDF-based context storage, context queries and subscription.

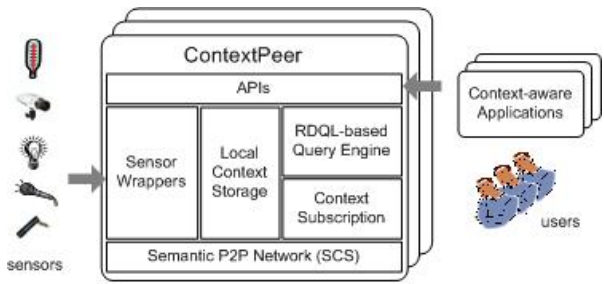

Figure 1: The architecture of ContextPeers

As shown in Figure 1, each ContextPeer consists of five components: the semantic P2P network layer, the sensor wrappers, the local context storage, the RDQL(RDF Data Query Language)-based query engine and the context subscription. The sensor wrappers capture various sensor data from physical or virtual sensors, convert them into RDF-based context data (i.e., in the form of RDF triples) and store the triples into the local context storage. A ContextPeer will then join the semantic P2P network based on the semantics of its local context data. The RDQL-based query engine parses and resolves context queries from users or applications. The context subscription registers subscription requests and notifies context consumers when context changes occur. The underlying SCS protocol layer is responsible for overlay construction and maintenance, and query routing. Application developers utilize a set of APIs provided by our system to access the functionalities of ContextPeers and build various context-aware applications in multiple smart spaces.

## 3. Context-aware Applications in Multiple Smart Spaces

Many existing or potential context-aware applications can use our system, especially those collaborative context-aware applications over multiple smart spaces such as Family Intercom [5]. For a family whose members live in different places, a reminder application may trigger an alert to all the members for the purpose of family activity planning or when a pre-set context arises. In health care domain, a tele-medical record application provides anywhere availability of a patient's medical history, or a tele-monitoring application tracks a patient wherever he/she goes.

To demonstrate how application developers could benefit from ContextPeers in real life, we are developing two typical collaborated context-aware applications: *Multi-Homes* and *Shopping-Assist*. The *Multi-Homes* application includes a set of applications that monitor home appliances and users, assist users' communication and provide intelligent services such as activity planning in multiple home environments. They are particularly useful for multiple families located in different geographic places. The *Shopping-Assist* application provides users helpful suggestions in making a buying decision based on what are needed from home and what are available in different shops. It accesses and utilizes cross-domain context information such as inventory context data in a smart home, user's location that changes from one location to another, merchandise data from various shops, etc.

## 4. Future Work and Conclusion

This paper presents a P2P context lookup system for multiple smart spaces. The system offers a uniform interface for acquiring context data from sensors, resolving context queries and notifying users and applications when context changes. We are currently developing cross-domain context-aware applications, and in the process of evaluating our system and applications in a real-life setting.